\documentclass[english,aps,manuscript,preprintnumbers]{revtex4-1}
\usepackage[T1]{fontenc}
\usepackage[latin9]{inputenc}
\usepackage{color}
\usepackage{babel}
\usepackage{amsmath}
\usepackage{amssymb}
\usepackage{graphicx}
\usepackage[unicode=true,pdfusetitle,
 bookmarks=true,bookmarksnumbered=false,bookmarksopen=false,
 breaklinks=false,pdfborder={0 0 1},backref=false,colorlinks=true]
 {hyperref}
\hypersetup{
 linkcolor=blue,citecolor=blue,urlcolor=blue}

\makeatletter
\newenvironment{lyxlist}[1]
{\begin{list}{}
{\settowidth{\labelwidth}{#1}
 \setlength{\leftmargin}{\labelwidth}
 \addtolength{\leftmargin}{\labelsep}
 }}
{\end{list}}

\makeatother

\begin{document}

\preprint{BROWN-HET-1725}

\title{BMS symmetry, soft particles and memory }

\author{Atreya Chatterjee}
\email{atreya_chatterjee@brown.edu}

\selectlanguage{english}%

\author{David A. Lowe}
\email{lowe@brown.edu}

\selectlanguage{english}%

\affiliation{Department of Physics, Brown University, Providence, Rhode Island
02912, USA}
\begin{abstract}
In this work, we revisit unitary irreducible representations of the
Bondi-Metzner-Sachs (BMS) group discovered by McCarthy. Representations
are labelled by an infinite number of super-momenta in addition to
four-momentum. Tensor products of these irreducible representations
lead to particle-like states dressed by soft gravitational modes.
Conservation of 4-momentum and supermomentum in the scattering of
such states leads to a memory effect encoded in the outgoing soft
modes. We note there exist irreducible representations corresponding
to soft states with strictly vanishing four-momentum, which may nevertheless
be produced by scattering of particle-like states. This fact has interesting
implications for the S-matrix in gravitational theories.
\end{abstract}
\maketitle
\tableofcontents{}

\section{Introduction}

One of the first breakthroughs in laying the foundation for an understanding
of holography in Minkowski space was the work of Bondi-Metzner-Sachs
\citep{Sachs103,10.2307/2414436}. It revealed that asymptotic symmetry
group of Minkowski space is a group of large diffeomorphisms called
the BMS group. Representations of the Poincare group \citep{10.2307/1968551}
have played an important role in classifying elementary particles
by their mass and spin. That motivates understanding the representations
of the BMS group and its connection to elementary particles. In the
1970s McCarthy studied the positive energy unitary irreducible representations
of BMS group \citep{McCarthy301,McCarthy317,McCarthy517,McCarthy489,PhysRevLett.32.565,doi:10.1063/1.1665917}.
But after this initial work, the subject has received little attention.
The physical interpretation of the representations was not entirely
clear at the time. In this work, we study from a physical viewpoint
most of the interesting representations with the aim of identifying
the interesting representations needed to construct a holographic
dual. These include massive and massless particles and also soft particles
with vanishing four-momentum. We find that in addition to zero momentum
limit of massless particles there are many new soft modes predicted
by BMS group which are related to gravitational memory \citep{zeldovich,braginsky}.

Recently, Strominger et. al. have discovered a relation between the
BMS group, soft theorems and the memory effect \citep{He:2014cra,He:2014laa,Strominger:2014pwa}.
They related supertranslations to memory effect \citep{Strominger:2014pwa}
which led them to propose that black hole carries soft hair \citep{PhysRevLett.116.231301}.
In this work we show that supertranslation charges indeed retain information
about the initial states via a straightforward group theory construction.
We consider a case where two particles collide and move away in different
directions. Conservation of momenta (including supermomenta) reveals
that final state has information about soft particles that stores
information about the initial state. Another interesting discussion
of the memory effect in electromagnetism appears in \citep{Susskind:2015hpa,Pasterski:2015zua}.

Other recent papers on the BMS group include a realization \citep{Longhi:1997zt,Gomis:2015ata}
on a scalar field, and more generally relation between the BMS group
and elementary particles\citep{Dappiaggi:2004cp}. The connection
between BMS group and non-relativistic conformal group, also known
as Galilean group \citep{Bagchi:2009pe,Bagchi:2010eg,Bagchi:2012cy,Bagchi:2016bcd}
has also been explored. The BMS charge algebra has been studied in
\citep{Barnich:2011mi} and BMS representations in three dimensions
have been explored in the following papers \citep{Barnich:2014kra,Barnich:2015uva,Campoleoni:2016vsh}.

In the present work we begin by reviewing the BMS group and establishing
notation. We then revisit some of the most relevant results from McCarthy's
classification of unitary irreducible representations of the BMS group
and connect the Bondi mass aspect to the function space on which BMS
is realized. We try to highlight only the physically important representations
and find all the massive and massless representations that appear
in the usual Wigner classification of the Poincare group, as well
as extra representations with differing supermomenta structures. The
group invariant norms associated with these families of representations
are constructed, which is an essential step in any attempt at capturing
the bulk dynamics via a holographic description. We then consider
tensor products/scattering of these states which allows us to explore
the extent to which gravitational memory allows the initial state
to be reconstructed from a final state. We conclude with some comments
on the relevance of the results to general gravitational S-matrix
theories in asymptotically flat spacetime such as string theory, and
the prospects for developing holographic models with BMS as a fundamental
symmetry group.

\section{Representations of the BMS group\label{sec:Little-groups-and}}

Asymptotic flatness requires that the Weyl tensor of the metric must
fall off like $O\left(r^{-3}\right)$ for large $r$ \citep{10.2307/2414436}
(for a recent review see \citep{Strominger:2017zoo}) which allows
the choice of the following asymptotically flat coordinates at leading
order at large $r$
\begin{eqnarray}
ds^{2} & = & -du^{2}-2dudr+2r^{2}\gamma_{z\bar{z}}dzd\bar{z}+2\frac{m_{B}(u,z,\bar{z})}{r}du^{2}+rC_{zz}dz^{2}\nonumber \\
 &  & +D^{z}C_{zz}dud\bar{z}+c.c+...\label{eq:ametric}
\end{eqnarray}
The function $m_{B}(u,z,\bar{z})$ is called the Bondi mass aspect
and the other coefficients are functions only of $u,z$ and $\bar{z}$.
The covariant derivative $D^{z}$ is defined with respect to the metric
on the unit sphere $\gamma_{z\bar{z}}=\frac{2}{(1+z\bar{z})^{2}}$.
In the next subsection we give a brief introduction to BMS group.
Then we show that the invariant mass function introduced in \citep{McCarthy489}
and the Bondi mass aspect are to be identified.

\subsection{BMS group\label{subsec:BMS-group}}

The group of diffeomorphisms which preserve the form of the metric
\eqref{eq:ametric} is called the BMS group. It is given by 
\begin{eqnarray*}
B & = & A\ltimes G
\end{eqnarray*}
where $G=SL(2,\mathbb{C})$ and $A$ is the abelian group of pointwise
addition of functions functions on a 2-sphere \citep{doi:10.1063/1.1665917}.
To make this statement well-defined we must specify more carefully
the class of functions to be considered. We follow the definition
of \citep{McCarthy489} and take these to be $C^{\infty}$ which implies
that the representation of $G$ on $A$ is equivalent to the operator
representation of $G$ on the space $D_{(2,2)}$ \citep{gelfandv5}.

The space $D_{(2,2)}$ is characterized by a pair of functions $\xi(z)$
and $\hat{\xi}(z)$ on the complex plane, which may be thought of
as functions on patches centered at the north and south poles of the
sphere respectively. These functions are $C^{\infty}$ everywhere
except at the origin and are related by the overlap condition
\[
\hat{\xi}(z)=|z|^{2}\xi(z^{-1})
\]
The action of $SL(2,\mathbb{C})$ element $\left[\begin{array}{cc}
\alpha & \beta\\
\gamma & \delta
\end{array}\right]$ is given by
\begin{eqnarray}
g\xi(z) & = & |\alpha+\gamma z|^{2}\xi\left(\frac{\beta+\delta z}{\alpha+\gamma z}\right)\label{eq:lorentz-2}\\
g\hat{\xi}(z) & = & |\beta+\delta z|^{2}\hat{\xi}\left(\frac{\alpha+\gamma z}{\beta+\delta z}\right)\nonumber 
\end{eqnarray}

We will be mostly interested in the dual space of $A$. As we will
see this corresponds most directly to the class of functions $m_{B}(u,z,\bar{z})$
that appear for some fixed value of $u$. The dual space corresponds
to the space $D_{(-2,-2)}$ in the notation of \citep{gelfandv5}
and, as we will see, is a space of distributions with a class of allowed
singularities. It is specified again by a pair of functions satisfying
the matching condition
\begin{eqnarray*}
\hat{\phi}(z) & = & |z|^{-6}\phi(z^{-1})
\end{eqnarray*}
The action of $G$ is given by
\begin{eqnarray}
g\phi(z) & = & |\alpha+\gamma z|^{-6}\phi\left(\frac{\beta+\delta z}{\alpha+\gamma z}\right)\label{eq:lorentz}\\
g\hat{\phi}(z) & = & |\beta+\delta z|^{-6}\hat{\phi}\left(\frac{\alpha+\gamma z}{\beta+\delta z}\right)\nonumber 
\end{eqnarray}
The 4-momentum associated with the functions $\phi(z)$ may be extracted
via the projector $\Pi$ expressed as the integral
\begin{align}
\Pi\phi(z') & =\frac{i}{\pi}\int dzd\bar{z}(z-z')(\bar{z}-\bar{z}')\phi(z)\label{eq:fourmomenta}\\
 & =\frac{i}{\pi}\left((p^{0}+p^{3})+(p^{0}-p^{3})z'\bar{z}'-(p^{1}-ip^{2})z'-(p^{1}+ip^{2})\bar{z}'\right)
\end{align}
which is a polynomial of weight 2 in $z'$, with coefficients corresponding
to the 4-momenta \citep{McCarthy489} $p^{\mu}$. For this to be well-defined,
the regulator as $|z|\to\infty$ implicit in the definition of the
$D_{(-2,-2)}$ distributions must be taken into account. This can
therefore be rewritten in terms of convergent integrals as
\begin{equation}
\Pi\phi(z')=\frac{i}{\pi}\int_{|z|<1}dzd\bar{z}\left((z-z')(\bar{z}-\bar{z}')\phi(z)+(1-zz')(1-\bar{z}\bar{z}')\hat{\phi}(z)\right)\label{eq:fourmomentumreg}
\end{equation}

The higher order terms in $\phi(z)$ are labelled by the supermomenta.
The supermomenta form a $G$ invariant subspace, implying that an
irreducible representation of the BMS group describes states with
the same mass (i.e. 4-momentum squared). Equation \eqref{eq:fourmomenta}
matches equation 72 in \citep{10.2307/2414436} which gives the Bondi
4-momentum in terms of an integral of the Bondi mass aspect $m_{B}(u,z,\bar{z})$.
Thus we may identify $\phi(z)$ with $m_{B}$ up to a rescaling factor,
and the derived 4-momenta behave as expected under $G$.

In turn, this provides a more physical justification for the choice
of the space of functions $D_{(-2,-2)}$. This space of distributions
yield 4-momenta corresponding to finite center of mass energies, as
well as finite supermomenta, and prescribed fall-off conditions \citep{gelfandv5}
that guarantee integrals such as \eqref{eq:fourmomenta} are well-defined.

\subsection{Little groups}

As with Wigner's classification of the irreducible representations
of the Poincare group, the first step in understanding representations
is to understand little groups. One may then construct the irreducible
representations via the method of induced representations \citep{10.2307/1968551,mackey1968induced},
lifting representations of the subgroup to representations of the
group.

In Wigner's classification, one identifies classes of four-momenta
invariant under Poincare subgroups. For BMS the goal is to find functions
$\phi(z)$ invariant under the little groups of BMS. McCarthy give
a detailed list of most of the little groups \citep{McCarthy489}.
Here we discuss some of them in detail. 

\subsubsection{$SU(2)$\label{subsec: su2}}

The first important little group is $SU(2).$ The class of functions
invariant under this group is
\begin{eqnarray}
\phi(z) & = & \hat{\phi}(z)=m\left(1+|z|^{2}\right)^{-3}\label{eq:su2func}
\end{eqnarray}
This represents a particle of mass $m$ at rest. One can check that
the 4-momentum $\left(p_{0},p_{1},p_{2},p_{3}\right)$ indeed transforms
correctly under the action of Lorentz generators \eqref{eq:lorentz}.
As an example, let us look at the action of boost $g_{t}=\left[\begin{array}{cc}
e^{t/2} & 0\\
0 & e^{-t/2}
\end{array}\right]$. Acting on \eqref{eq:su2func}
\begin{eqnarray}
g\phi(z) & = & me^{-3t}\left(1+|z|^{2}e^{-2t}\right)^{-3}=m\left(e^{-t}|z|^{2}+e^{t}\right)^{-3}\label{eq:boostm}
\end{eqnarray}
 Using equation \eqref{eq:fourmomenta} we get
\begin{eqnarray*}
\Pi\phi(z') & = & \frac{i}{\pi}\int dzd\bar{z}(z-z')(\bar{z}-\bar{z}')m\left(e^{-t}|z|^{2}+e^{t}\right)^{-3}=m\left(e^{t}+e^{-t}|z|^{2}\right)
\end{eqnarray*}
which leads via \eqref{eq:fourmomenta} to $p^{0}=m\cosh t$ and $p^{3}=m\sinh t$
as expected for a boost. 

Also note that the super-momenta get populated by the action of the
boost due to the higher order terms present in \eqref{eq:boostm}
beyond order 2. Thus the simplest representation of BMS is that of
a massive particle, matching what one expects of the Poincare group,
but the representation traces out an orbit in the infinite dimensional
space of supermomenta as one acts with Lorentz generators.

The representations of the little group may also carry spin $\ell$
which is half-integer. As shown in \citep{McCarthy301} this yields
a single spin $\ell$ representation of the Poincare subgroup of BMS.

\subsubsection{$\Delta$\label{subsec:delta}}

The second important little group is $\Delta$, in the notation of
\citep{McCarthy489}, or more commonly the Euclidean group in two
dimensions $E(2)$. It yields usual massless particles, and as above,
Lorentz transformation fill out an orbit in supermomentum space. This
corresponds to the invariant functions
\begin{align*}
\phi(z) & =K\\
\hat{\phi}(z) & =K|z|^{-6}+A\frac{\partial^{2}}{\partial z^{2}}\frac{\partial^{2}}{\partial\bar{z}^{2}}\delta(z)+B\frac{\partial^{2}}{\partial z^{2}}\delta(z)+\bar{B}\frac{\partial^{2}}{\partial\bar{z}^{2}}\delta(z)+C\delta(z)
\end{align*}
Note here $\delta(z)\equiv\delta(\mathrm{Re}z)\delta(\mathrm{Im}z)$,
and likewise we suppress the $\bar{z}$ dependence of $\phi$,$\hat{\phi}$.
Here $A$ and $C$ are real, and $B$ is complex. This clearly illustrates
the need for the $D_{(-2,-2)}$ space of generalized functions to
correctly accommodate massless particles. These representations were
not present in the earlier studies \citep{McCarthy517,McCarthy317,McCarthy301}.
To evaluate four momentum on such a representation one must use the
formula \eqref{eq:fourmomentumreg} to properly regulate the otherwise
divergent expression \eqref{eq:fourmomenta}. Finite 4-momenta are
obtained provided $K=0$. In this case, $C$ is proportional to the
light-like 4-momentum.

The spin of these representations has been studied in \citep{McCarthy489}
and as expected one gets either a chiral massless representation with
a single Poincare spin $s=0,1/2,\cdots$. Alternatively one may get
one of the massless continuous spin representations of Wigner's classification,
whose physical significance remains unclear.

\subsubsection{$SL(2,C)$ }

In general one may take the entire group of Lorentz transformations
to be a little group, in which case the invariant functions take the
form
\begin{align*}
\phi(z) & =\hat{\phi}(z)=0
\end{align*}
which implies vanishing of the 4-momentum and of all the supermomentum.
Nevertheless, one may pick a unitary representation of the little
group and lift it to a representation of BMS. It is natural to think
of such representations as arising from a unitary irreducible representation
corresponding to a massive (or massless) field on an internal three-dimensional
de Sitter spacetime $dS_{3}$ \citep{Bekaert:2006py}. Such representations
are infinite-dimensional. In any case, the situation here is unchanged
from the usual Poincare group. The standard procedure is to throw
out all but the trivial representation, leaving the Poincare invariant
vacuum as the unique state with vanishing 4-momentum. Lifting to BMS,
we obtain a unique state with vanishing 4-momentum and supermomentum.
Since the other infinite-dimensional families of states are not generated
from tensor products of the other states we will consider with the
vacuum, we can safely ignore these exotic infinite dimensional representations
with vanishing momentum. 

\subsubsection{$SL(2,R)$}

The situation is more interesting for this maximal little group. In
this case the invariant functions take the form
\begin{align*}
\phi(z) & =K\left(\frac{z-\bar{z}}{i}\right)^{-3}+A\delta^{2}\left(\frac{z-\bar{z}}{i}\right)\\
\hat{\phi}(z) & =K\left(\frac{\bar{z}-z}{i}\right)^{-3}+A\delta^{2}\left(\frac{\bar{z}-z}{i}\right)
\end{align*}
where $K$ and $A$ are real parameters. For the Poincare group, this
little group would usually give rise to the tachyonic representations
where $p_{\mu}p^{\mu}<0$. Here the nuclear topology restricts the
class of distributions to those with vanishing 4-momentum when inserted
into \eqref{eq:fourmomentumreg}. Nevertheless, the higher order terms
present in the invariant functions generate a nontrivial orbit corresponding
to nonvanishing supermomentum.

As with the case of $SL(2,C)$ one can assign such representations
a nontrivial representation of the little group. In this case it would
correspond to a massive or massless field on an internal two-dimensional
de Sitter spacetime, which has the isometry group $SL(2,R)$. However
again such representations are infinite dimensional, and will not
arise from tensor products of the elementary representations we will
consider. These representations arise already in Wigner's classification
of the representations of the Poincare group, and are likewise not
thought to be physically relevant, because one can construct self-consistent
theories where they do not appear.

An exception is the trivial representation of the little group $SL(2,R)$.
Under the usual Poincare classification, these would be invariant
under a larger little group $SL(2,C)$ and so would be equivalent
to the $SL(2,C)$ invariant vacuum state. However under BMS such modes
carry nontrivial supermomentum. This leads to a class of ``soft modes''
which in general will be produced in the scattering of particle-like
states, and are in general necessary to enforce conservation of supermomentum.

\subsubsection{$\Gamma$}

For the Poincare group, the maximal little groups exhaust the set
of little groups. However for BMS it is also necessary to consider
the group $\Gamma$ which is a subgroup of all the above little groups
corresponding to rotations in a plane $\left(\begin{array}{cc}
\omega & 0\\
0 & \bar{\omega}
\end{array}\right)$ with $\omega$ a complex number of unit modulus. While the 4-momenta
invariant with respect to this little group are actually invariant
under a larger little group, this is no longer the case when the supermomenta
are included. The invariant function takes the form
\begin{align}
\phi(z) & =\beta(r)\nonumber \\
\hat{\phi}(z) & =r^{-6}\beta(1/r)\label{eq:gammaform}
\end{align}
where $z=re^{i\phi}$ with $\phi=[0,2\pi)$ and $r\geq0$. Here \textbf{$\beta$}
is a distribution satisfying the conditions above. The 4-momenta corresponding
to these representations may have $m^{2}=0$, $m^{2}>0$ or $m^{2}<0$.

For $m^{2}>0$ the representation corresponds \citep{McCarthy301}
to an infinite tower of Poincare spins labelled by some integer/half-integer
$j$ with the tower corresponding to all spins $s=j,j+1,\cdots$.

For $m^{2}=0$ and $m^{2}<0$ the Poincare representations are more
exotic, with integrals over continuous spins needed to generate the
BMS representation. 

\subsubsection{Indecomposable}

In the present work we are restricting our consideration to unitary
irreducible representations of the BMS group. It is possible this
is too restrictive a class of representations to build a useful holographic
description of asymptotically flat space. Because the BMS group is
non-compact, representations that may be decomposed into irreducible
representations are actually rather special, and more generally one
should consider indecomposable representations. As far as we are aware,
the classification of such representations for non-compact groups
is still relatively undeveloped.

\subsubsection{Non-connected subgroups}

There are a variety of non-connected little groups that can appear
as subgroups of the BMS group \citep{McCarthy489}. For simplicity
we do not consider these in the present work.

\section{Holography\label{sec:Invariant-norm}}

One of the main motivations for considering the irreducible representations
of the BMS group, is to get a better understanding of the basic ingredients
needed to build a holographic description of the theory on null infinity
$\mathcal{I}$. The same considerations also apply when considering
the allowed set of asymptotic states in an S-matrix description of
a gravitational theory. As such, we now turn our attention to defining
BMS invariant norms for the representations of interest, and see that
these may be realized as integrals on $\mathcal{I}$. 

In the general case the norm is defined using the group invariant
measure on the coset space $G/H$ where $G=SL(2,C)$ and $H$ is the
little group \citep{McCarthy517}
\begin{eqnarray*}
\int f(g)d\mu(g) & = & \int_{G/H}\left(\int_{H}f(gh)d\mu(h)\right)d\mu_{G/H}\,.
\end{eqnarray*}

\subsection{$SU(2)$ and $\Delta$}

It is perhaps simplest to begin in momentum space. As we have seen
for the $SU(2)$ little group, we have representations of BMS that
essentially coincide with ordinary massive particle representations
of the Poincare group. The same is true for massless particles and
the little group $\Delta$. Wigner has given the invariant norm for
these two subgroups as
\begin{eqnarray*}
\left(\psi_{1},\psi_{2}\right) & = & \int_{0}^{\infty}\psi_{1}(p)\psi_{2}(p)\frac{dp_{1}dp_{2}dp_{3}}{p_{4}}\,.
\end{eqnarray*}
As we see, this integral may be viewed as an on-shell integral in
the bulk $p_{4}^{2}=m^{2}+\sum_{i}p_{i}^{2}$, or as an off-shell
integral over the holographic boundary $\mathcal{I}$.

\subsection{$SL(2,R)$}

Again the little group is three-dimensional, but now the invariant
norm can be interpreted as an integral over three-dimensional de Sitter
spacetime which corresponds to the coset $SL(2,C)/SL(2,R)$
\[
(\psi_{1},\psi_{2})=\int_{\sum_{_{i}}p_{i}^{2}=1}^{\infty}\psi_{1}(p)\psi_{2}(p)\frac{dp_{1}dp_{2}dp_{3}}{\sqrt{\sum_{i}p_{i}^{2}-1}}
\]

\subsection{$\Gamma$}

Since $\Gamma$ is only one-dimensional the coset space will be five-dimensional
and may be written as an integral over on-shell 4-momenta ($p_{4}^{2}=\sum_{i}p_{i}^{2}+m^{2})$
supplemented by a pair of angles
\begin{eqnarray}
\left(\psi_{1},\psi_{2}\right) & = & \int_{0}^{\infty}\psi_{1}(p,\theta)\psi_{2}(p,\theta)\frac{dp_{1}dp_{2}dp_{3}d\theta_{1}d\theta_{2}}{p_{4}}\label{eq:gammanorm}
\end{eqnarray}
which may be interpreted as an integral over $\mathcal{I}$ and two
internal degrees of freedom $\theta$.

\section{Scattering examples}

By taking tensor products of the above representations of BMS we can
gain insight into how the symmetry constrains the scattering of particle-like
representations and study what BMS representations appear when ordinary
particles undergo scattering.

\subsection{Particles forming bound state\label{subsec:Anti-linear-particles-forming}}

\begin{figure}
\begin{lyxlist}{00.00.0000}
\item [{\centering}] \includegraphics[bb=0bp 0bp 371bp 368bp,scale=0.6]{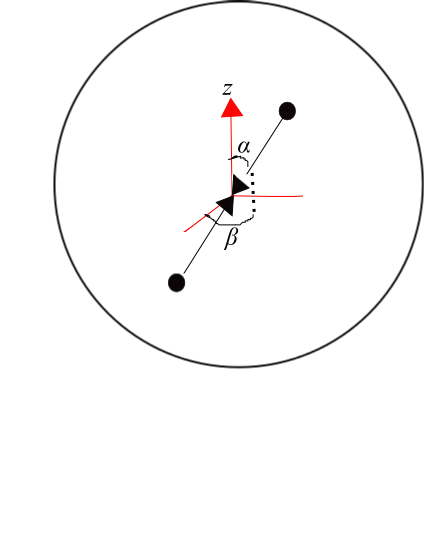}
\end{lyxlist}
\caption{\label{fig:The-figure-depicts}The figure depicts two particles of
mass $M$ moving in opposite directions forming a bound state.}
\end{figure}

Consider a representations of the $SU(2)$ little group corresponding
to two particles with mass $M$. One is moving in angular direction
$(\theta,\phi)=\left(\alpha,\beta\right)$ and the other in the opposite
direction $\left(\alpha-\pi,\beta\right)$. After sometime they collide
and form a bound state as shown in figure\eqref{fig:The-figure-depicts}
The mass aspect functions of a particle can be found by boosting \eqref{eq:su2func}
along $z$-axis and then rotating by $\alpha$ around $y$-axis followed
by rotation around $z$-axis by $\beta$. That is
\begin{eqnarray*}
\phi_{+\alpha} & = & g_{\alpha}\phi_{z+}
\end{eqnarray*}
where
\begin{eqnarray*}
\phi_{+\alpha}(\theta,\phi) & = & \left[\begin{array}{cc}
\cos\frac{\alpha}{2} & \sin\frac{\alpha}{2}\\
-\sin\frac{\alpha}{2} & \cos\frac{\alpha}{2}
\end{array}\right]\phi_{z+}=\phi_{z+}\left(\frac{z\cos\frac{\alpha}{2}-\sin\frac{\alpha}{2}}{z\sin\frac{\alpha}{2}+\cos\frac{\alpha}{2}}\right)=\left(\frac{e^{t}|\frac{z\cos\frac{\alpha}{2}-\sin\frac{\alpha}{2}}{z\sin\frac{\alpha}{2}+\cos\frac{\alpha}{2}}|^{2}+e^{-t}}{1+|\frac{z\cos\frac{\alpha}{2}-\sin\frac{\alpha}{2}}{z\sin\frac{\alpha}{2}+\cos\frac{\alpha}{2}}|^{2}}\right)^{-3}M\\
 &  & =\frac{M}{\left(\cosh t-\sinh t\cos\theta\cos\alpha-\sinh t\sin\theta\cos\phi\sin\alpha\right)^{3}}
\end{eqnarray*}
and rotating by $\beta$ around $z$-axis gives
\begin{eqnarray*}
\phi_{+\alpha,\beta}(\theta,\phi) & = & \frac{M}{\left(\cosh t-\sinh t\cos\theta\cos\alpha-\sinh t\sin\theta\cos\left(\phi+\beta\right)\sin\alpha\right)^{3}}
\end{eqnarray*}
Now consider a particle moving in opposite direction. That is angular
coordinates $\left(\alpha-\pi,\beta\right)$. 
\begin{eqnarray*}
\phi_{-\alpha,\beta}(\theta,\phi) & = & \frac{M}{\left(\cosh t+\sinh t\cos\theta\cos\alpha+\sinh t\sin\theta\cos\left(\phi+\beta\right)\sin\alpha\right)^{3}}
\end{eqnarray*}
At the linearized level, the mass aspect function of the whole system
is 
\begin{eqnarray}
\phi_{\alpha,\beta}(\theta,\phi) & = & \phi_{+\alpha,\beta}(\theta,\phi)+\phi_{-\alpha,\beta}(\theta,\phi)\nonumber \\
 & = & \frac{M}{\left(\cosh t-\sinh t\cos\theta\cos\alpha-\sinh t\sin\theta\cos\left(\phi+\beta\right)\sin\alpha\right)^{3}}\\
 &  & +\frac{M}{\left(\cosh t+\sinh t\cos\theta\cos\alpha+\sinh t\sin\theta\cos\left(\phi+\beta\right)\sin\alpha\right)^{3}}\label{eq:phialphabeta}
\end{eqnarray}
Since two particles are moving in opposite directions, in no frame
will both the particles be at rest together. Neither boosts nor rotations
can transform the above function into a constant. The 4-momenta may
be evaluated using \eqref{eq:fourmomenta} and are given by $p_{0}=2M\cosh t,p_{1}=p_{2}=p_{3}=0.$
The higher momentums corresponding to supermomenta are nontrivial,
and are functions of $\alpha,\beta$. Performing a rotation of $-\beta$
around $z$-axis followed by $-\alpha$ around $y$-axis transforms
\eqref{eq:phialphabeta} to a function of $\cos\theta$ only. This
implies the function is invariant under $\Gamma$ little group and
no bigger subgroup of Lorentz group. 

This construction also provides insight into the invariant norm for
the $\Gamma$ representations \eqref{eq:gammanorm}. While boosts
fill out three dimensions of the associated states as usual, one needs
an extra integral over the angular directions corresponding to $(\alpha,\beta)$
to generate the complete set of associated states, yielding the five-dimensional
integral in \eqref{eq:gammanorm}. 

So we come to an interesting conclusion. The mass aspect functions
of $\Gamma$ can be viewed as tensor product of reps of $SU(2)$.
One may perform essentially the same computation for the massless
representations associated with the little group $\Delta$ replacing
those of $SU(2)$. BMS representation of the final system retains
memory about the direction of the incoming particles. In this case
of two-body scattering, the supermomenta allow all the information
about the initial state of the system to be retrieved from the final
bound state. This is in line with the soft hair proposal of Strominger
et al \citep{PhysRevLett.116.231301}.

\subsection{Soft modes\label{subsec:Soft-particle-in} in scattering}

\begin{figure}
\begin{lyxlist}{00.00.0000}
\item [{\centering}] \includegraphics[bb=0bp 0bp 371bp 368bp,scale=0.6]{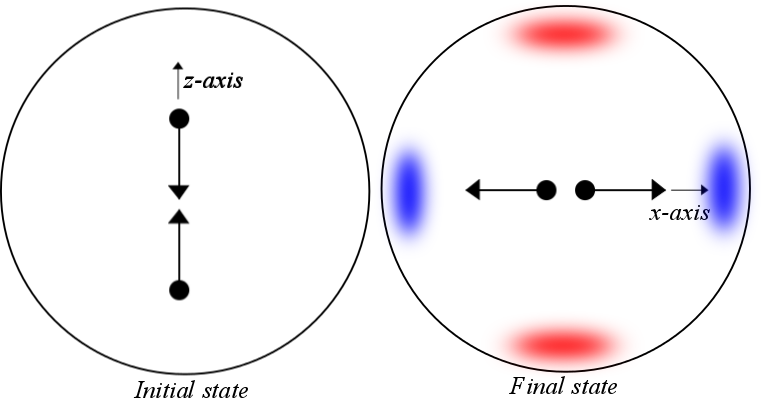}
\end{lyxlist}
\caption{\label{fig:The-left-figure}The left figure represent two particle
of mass $m$ moving along $z$-axis. They collide and move out along
$x$-axis. The figure on the right represents the final state. Subtracting
blue patch from the red patch on the celestial sphere gives the soft
mode in the final state. }
\end{figure}

Extending the above considerations, we now consider $2\to2$ scattering.
Consider an initial state is $\phi_{z}$ and a final state $\phi_{x}$
accompanied by soft modes. Figure \eqref{fig:The-left-figure} shows
the process.
\begin{eqnarray*}
\phi_{initial}(\theta,\phi)=\phi_{z}(\theta,\phi) & = & (2M,0,0,0,\{p_{l,m}\}_{z})=\frac{M}{\left(\cosh t-\sinh t\cos\theta\right)^{3}}+\frac{M}{\left(\cosh t+\sinh t\cos\theta\right)^{3}}\in\Gamma
\end{eqnarray*}
Part of the final state is two particles going along the $x$-axis
\begin{eqnarray*}
\phi_{x}(\theta,\phi) & = & \frac{M}{\left(\cosh t-\sinh t\sin\theta\cos\phi\right)^{3}}+\frac{M}{\left(\cosh t+\sinh t\sin\theta\cos\phi\right)^{3}}
\end{eqnarray*}
By conservation of supermomenta, the initial mass aspect should match
the final mass aspect
\begin{eqnarray*}
\phi_{initial} & = & \phi_{final}\\
\phi_{z}(\theta,\phi) & = & \phi_{x}(\theta,\phi)+\phi_{soft}\\
(2M,0,0,0,\{p_{l,m}\}_{z}) & = & (2M,0,0,0,\{p_{lm}\}_{x})+(0,0,0,0,\{p_{lm}\}_{z}-\{p_{lm}\}_{x})
\end{eqnarray*}
In this case, while the outgoing massive particles transform under
the standard $SU(2)$ little groups, there is an additional soft mode
with vanishing 4-momentum but non\textendash vanishing supermomentum.
In this case the soft mode transforms under the $\Gamma$ little group
and represents the gravitational memory effect. 

\section{Conclusion}

Many of the results we have discussed appear in McCarthy's original
works but have been passed over in much of the subsequent literature,
and our goal was to cast the most relevant selection of these results
in a modern context, where they may be of use to researchers attempting
holographic formulations of asymptotically flat spacetime, or simply
trying to understand gravitational memory from the perspective of
the BMS group. We started with a brief introduction to the BMS group
and identified 4-momenta and the supermomenta. Representations of
$\Delta,SU(2)$ represent massless and massive particles respectively
corresponding directly to Wigner's original classification of the
Poincare group. Then we derive the invariant measure and invariant
norm for some of the little groups. This revealed that invariant norm
of little groups other than $SU(2),\Delta$ involves integrating over
a larger phase space. Specifically for $\Gamma$ one encounters integrals
over 5 dimensions. Starting with a representative state of $\Gamma$,
both rotation and boosts are required to traverse complete orbit inside
$\Gamma$. This implies that rotations produce states which cannot
be obtained just by boosts. This is related to the fact that representations
of $\Gamma$ can be expressed as bound state of rep of $SU(2),\Delta$.
To explore this point we considered two particles moving in opposite
directions forming a bound state. Momenta of final state depend on
the direction of initial particles. In other words, BMS representations
store not just the total 4-momenta of the system but also retain information
about the individual 4-momenta of the initial state. This is in contrast
to Poincare representations where the final state just depends on
total energy. 

These results should have important implications for any $S-$matrix
theory of gravity in asymptotically flat spacetime. In string theory,
for example, these $S-$matrix elements are built using vertex operators
corresponding to representations of the Poincare group. For such a
description to be consistent it is implicit that the scattering states
of such particles form a complete set. According to our analysis of
the BMS group, that is not the case. For example, there exist unitary
irreducible representations of the BMS group with vanishing 4-momenta
but non-vanishing supermomenta that are not limits of massless particles
(with non-vanishing light-like 4-momentum) such as the soft mode representations
of the $SL(2,R)$ little group that we discussed. One also has irreducible
representations of the little group $\Gamma$ that can also generate
soft modes with vanishing 4-momenta, but non-vanishing supermomenta.
On the other hand, it is clear there is a unique vacuum state, the
trivial representation of the BMS group, which is of course invariant
under all the asymptotic symmetries. There has been some preliminary
discussion of some of these issues in the bosonic string \citep{Avery:2015gxa}
but we believe the present results warrant further study of the spectrum
of string theory to obtain a more complete understanding of the soft
modes. 

From the perspective of holography the present work shows what irreducible
representations of the BMS group are needed to formulate the elementary
ingredients of such a description. There is some commonality with
the AdS/CFT approach, namely a holographic ``operator'' transforming
as an irrep of BMS in one-to-one correspondence with bulk fields with
fixed mass and spin. Such operators naturally live in a three-dimensional
space according to the norms described in section \ref{sec:Invariant-norm}.
However the existence of the more exotic representations discussed
above suggest this picture in not complete in the case of BMS. For
example if representations of the little group $\Gamma$ must be introduced
as elementary operators in the holographic description, they naturally
live in a five-dimensional space. Furthermore the operators corresponding
to the $SL(2,R)$ representations will serve to generate states with
nontrivial supermomenta, with no cost in 4-momentum. These representations
appear to live in an auxiliary three-dimensional de Sitter spacetime.
From the usual perspective, this would imply the vacuum is highly
degenerate, making it difficult to construct a reasonable interacting
theory based on such operators at the quantum level. In any case,
we hope the present work goes some way to highlighting the obstacles
that need to be addressed in formulating holography in asymptotically
flat spacetime.
\begin{acknowledgments}
This research is supported in part by DOE grant de-sc0010010.
\end{acknowledgments}

\bibliographystyle{/Users/lowe/Desktop/current/utphys}
\bibliography{reps_of_BMS}

\end{document}